\documentstyle[aps,prl,floats,epsfig]{revtex}

\newcommand{\ket}[1]{|#1\rangle}
\newcommand{\bra}[1]{ \langle #1 \,  |}
\newcommand{\eins}{\mbox{$1 \hspace{-1.0mm}  {\bf l}$}}
\begin{document}

\title{Optimal estimation of multiple phases}

\author{Chiara Macchiavello}
\address{Dipartimento di Fisica ``A.Volta", Via Bassi 6,
I-27100 Pavia, Italy \\ 
and Istituto Nazionale per la Fisica della Materia (INFM)}

\maketitle

\begin{abstract}
We study the issue of simultaneous estimation of several phase shifts 
induced by commuting operators on a quantum state. 
We derive the optimal positive operator-valued measure 
corresponding to the multiple-phase estimation.  
In particular, we discuss the explicit case of the optimal 
detection of double phase for a system of identical qutrits and generalise
these results to optimal multiple phase detection for $d$-dimensional quantum
states. 
\end{abstract}

\pacs{03.65.-w, 03.67.-a}

\section{Introduction}
\label{s:intro}

The issue of phase estimation has important applications in
quantum computation and quantum information theory. For example, it was shown
that the existing quantum algorithms can be described in a unified way
as quantum interference  processes among different computational paths where 
the result of the computation is encoded in a phase shift \cite{cemm}.
The design of optimal phase measurement procedures is also crucial in 
various tasks of atomic physics, such as for example methods for precision 
spectroscopy \cite{fs}, and quantum interferometric experiments in
quantum optics.

The problem of the optimal estimation of the value of a phase shift experienced
by a quantum state has been extensively studied \cite{physcri}, and in
particular a method to derive the optimal measurement procedure in the
phase covariant case was reported in Ref. \cite{Holevo}. More recently, 
a general formulation of the phase estimation problem was considered in
Ref. \cite{pomph}, where a method to derive the optimal positive-operator
valued measurement (POVM) \cite{helstrom}
for a generally degenerate phase shift operator
was developed. In this work we introduce the problem of simultaneous estimation
of several phase shifts undergone by a quantum physical system. 
More specifically, we address the problem of estimating the values of 
$M$ independent phase-shifts 
$\phi_j$ ($j=1,M$), pertaining to the unitary transformation
\begin{eqnarray}
\rho_{\{\phi_j\}} =e^{-i\sum_{j=1}^M \phi_j \hat H_j}\,\rho_0\,e^{i\sum_{j=1}^M
\phi_j \hat H_j} \;
\label{unit}
\end{eqnarray}
where $\hat H_j$ represent $M$ commuting self-adjoint operators, which are 
in general degenerate on the Hilbert space ${\cal H}$ of the considered 
quantum system and each of them has a
discrete spectrum $S_j$ ($S_j$ can be for example ${\mbox{Z}}$, 
${\mbox{N}}$, or ${\mbox{Z}_q}$, $q>0$) \cite{nota}. In Eq. (\ref{unit}) 
$\rho_0$ is a generic initial pure state $\ket{\psi_0}\bra{\psi_0}$
describing a quantum system with arbitrary dimension. 
We want to point out that the 
scenario of simultaneous estimation of several phases may be useful to
improve the efficiency of quantum information processing tasks where several
variables are encoded into phases in the same quantum states. 

The paper is organised as follows. In Sect. \ref{s:problem} we 
derive a general treatment of the multiple phase estimation problem,
extending to the multi-phase case the approach presented in \cite{pomph} 
for the case of single phase estimation.
In Sect. \ref{s:qutrits} we derive the optimal POVM and the corresponding 
estimation fidelity for a system of $N$ identically prepared ``equatorial'' 
three-dimensional systems. 
We want to point out that the possibility of 
encoding information in the states of three dimensional systems has been the 
object of several recent studies, for example in the context of quantum cloning
\cite{clon3} and quantum cryptography \cite{qc3}. 
In Sect. \ref{s:qudits} we extend these results derived for qutrits
to the case of quantum systems with arbitrary finite dimension $d$.
Finally, we summarise and comment the results presented in Sect. \ref{s:conc}.

\section{Optimal POVM for multiple-phase estimation}
\label{s:problem}

In this section we derive the optimal POVM corresponding to the simultaneous 
estimation of several phase shifts experienced by a pure state 
$\ket{\psi_0}$, belonging to the Hilbert space ${\cal H}$, that undergoes
the unitary transformation (\ref{unit}).
Following the approach of Ref. \cite{pomph},
we treat the estimation problem in the general framework of quantum
estimation theory \cite{helstrom}. According to this framework, we first 
define a cost function 
$C(\{\bar\phi_j\},\{\phi_j\})$ which depends on the set of the $M$ estimated 
values $\{\bar\phi_j\}$, that are the results of the estimation procedure,
and on the set of the $M$ true values $\{\phi_j\}$. This function
weights the errors for the 
estimates $\{\bar\phi_j\}$ given the true values $\{\phi_j\}$.  
The estimation problem then consists in minimizing the 
average cost $\bar C$ of the procedure defined as
\begin{eqnarray}
\bar C=\int_0^{2\pi}d\phi_1\int_0^{2\pi}d\phi_2..\int_0^{2\pi}d\phi_M
\,p_0(\{\phi_j\})\int_0^{2\pi}d\bar\phi_1\int_0^{2\pi}d\bar\phi_2..
\int_0^{2\pi}d\bar\phi_M C(\{\bar\phi_j\},\{\phi_j\})
\,p(\{\bar\phi_j\}|\{\phi_j\})\;,
\label{avc}
\end{eqnarray}
where $p_0(\{\phi_j\})$ is the {\it a priori} probability density for the
true values $\{\phi_j\}$ and
$p(\{\bar\phi_j\}|\{\phi_j\})$ is the conditional probability of 
estimating the set of values $\{\bar\phi_j\}$ given the true values 
$\{\phi_j\}$.  
The average cost is minimized by optimizing the POVM
$d\mu(\{\bar\phi_j\})$ which appears in the definition of the conditional 
probability as follows
\begin{eqnarray}
p(\{\bar\phi_j\}|\{\phi_j\})d\bar\phi_1d\bar\phi_2...d\bar\phi_M
=\mbox{Tr}[d\mu(\{\bar\phi_j\})e^{-i\sum_{j=1}^M\phi_j \hat H_j}\rho_0 
e^{i\sum_{j=1}^M\phi_j \hat H_j}]\;.
\label{prob}
\end{eqnarray}

In this work we consider the general scenario where all the 
values $\{\phi_j\}$ are 
{\it a priori} uniformly distributed, i.e. the probability density is
simply given by
$p_0(\{\phi_j\})=(1/2\pi)^M$.  Moreover, we consider the case where 
the errors in the estimates are weighted independently
of the values $\phi_j$ of the phases, but they depend only on the values of
$\bar\phi_j-\phi_j$, so that the cost function becomes an even
function of $M$  variables, i.e.  $C(\{\bar\phi_j\},\{\phi_j\})\equiv
C(\{\bar\phi_j-\phi_j\})$.  From these requirements it 
follows that also the conditional
probability corresponding to the optimal estimation procedure
will depend only on the variables $\bar\phi_j-\phi_j$, 
and therefore the optimal POVM will be phase-covariant, i.e. of the form
\begin{eqnarray}
d\mu(\{\bar\phi_j\})=
e^{-i\sum_{j=1}^M\bar\phi_j \hat H_j}\chi 
e^{i\sum_{j=1}^M\bar\phi_j \hat H_j}\frac{d\bar\phi_1}{2\pi}
\frac{d\bar\phi_2}{2\pi}...\frac{d\bar\phi_M}{2\pi}\;.
\label{dmu}
\end{eqnarray}
In the above equation  $\chi$ is a positive operator 
satisfying the completeness constraints
needed for the normalization of
the POVM $\int d\mu(\{\phi_j\})=\eins$, where $\eins$ denotes the identity 
operator.  
Actually, using Eq. (\ref{prob}) and the invariance of the trace under
cyclic permutations it can be shown that
$p(\{\bar\phi_j\}|\{\phi_j\})\equiv p(\{\bar\phi_j-\phi_j\})$ if and only 
if $d\mu(\{\bar\phi_j\})$ is
covariant.  Therefore the optimization of the phase estimation procedure 
can be performed  by
finding the positive operator $\chi$ that minimises the average cost
for a given cost function $C(\{\phi_j\})$ and a
generic initial state $\rho_0$.  

We will now show explicitly how to derive the optimal POVM
for a broad class of cost functions and initial states $\rho_0$. 
First of all we will choose the
representation where all the operators $\hat H_j$ are diagonal. 
We have assumed 
that the operators $\hat H_j$ commute, so we can identify a common basis
of eigenvectors. 
The operators $\hat H_j$ are generally degenerate, 
and we will denote by $\ket{\{n_j\}}_\nu$ a choice of
(normalized) eigenvectors corresponding to eigenvalue $n_j$ for the operator
$\hat H_j$,  by $\Pi_{\{n_j\}}$ 
the projector onto the corresponding degenerate eigenspace and by $\nu$ a 
degeneracy index, whose maximum value corresponds to the dimension of the 
degenerate eigenspace. 

We will now generalise the projection method developed in \cite{pomph}
and define ${\cal H}_\parallel$ as the Hilbert space spanned by the
(normalized) vectors $\ket{\{n_j\}}\propto\Pi_{\{n_j\}}\ket{\psi_0}\neq 0$ 
with the
choice of the arbitrary phases such that $\langle \{n_j\}\ket{\psi_0}>0$.
We can then write the Hilbert space of the system as ${\cal H}=
{\cal H_\parallel}\otimes{\cal H_\perp}$, where the component
${\cal H_\perp}$ is spanned by states that are orthogonal to $\ket{\psi_0}$. 
Hence the POVM can be chosen of the block diagonal form on 
${\cal H_\parallel}\otimes{\cal H_\perp}$,
i.e. $d\mu(\{\phi_j\})=d\mu_\parallel (\{\phi_j\})
\oplus d\mu_\perp(\{\phi_j\})$. In this way the component
$d\mu_\perp(\{\phi_j\})$ of the POVM acting on ${\cal H_\perp}$
can be chosen arbitrarily because it does not contribute to the average cost.
Therefore, the optimisation of the estimation procedure can be performed by 
optimising only the component $d\mu_\parallel (\{\phi_j\})$ of the POVM.

In order to optimise the POVM we can assume that 
$\Pi_{\{n_j\}}\ket{\psi_0}\neq 0$ for all 
values of $\{n_j\}$, since the resulting POVM will be optimal 
also for states having zero projection for some of these values.  
Due to the covariance property (\ref{dmu}) and to the argument followed above,
we can also write $\chi=\chi_\parallel \oplus \chi_\perp$.
Thus, the problem reduces to finding the positive operator $\chi_\parallel$ 
that minimizes the cost $\bar C$ in Eq. (\ref{avc}).  
In order to accomplish this task, 
we first rewrite Eq. (\ref{avc}) more conveniently as

\begin{eqnarray}
\bar C=\int_0^{2\pi}\frac{d\phi_1}{2\pi}\int_0^{2\pi}\frac{d\phi_2}{2\pi}..
\int_0^{2\pi}\frac{d\phi_M}{2\pi}C(\{\phi_j\})\mbox{Tr}[\chi
e^{-i\sum_{j=1}^M\phi_j \hat H_j}\rho_0 e^{i\sum_{j=1}^M\phi_j \hat H_j}]\;.
\label{avc2}
\end{eqnarray}

We will now express the operator $\chi_\parallel$ on the $\ket{\{n_j\}}$ basis 
as follows
\begin{eqnarray}
\chi_\parallel=\sum_{\{n_j\},\{m_j\}}\ket{\{n_j\}}\langle \{m_j\}|
\chi_{\{n_j\}\{m_j\}}\;.
\end{eqnarray}

The positivity condition for the operator $\chi$ implies the inequalities
\begin{eqnarray}
|\chi_{\{n_j\}\{m_j\}}|\leq\sqrt{\chi_{\{n_j\}\{n_j\}}
\chi_{\{m_j\}\{m_j\}}}=1\;,
\label{mat}
\end{eqnarray}
where the last equality comes from the POVM completeness relation $\int
d\mu_\parallel (\phi)= \eins_{{\parallel}}$. 

The cost functions we will consider are
$2\pi$-periodic functions  in the variables $\{\phi_j\}$, and therefore
they can be written as 
\begin{equation}
C(\{\phi_j\})=-\sum_{l_1,l_2,...l_M=-\infty}^{\infty}
c_{\{l_j\}}e^{i\sum_jl_j\phi_j}\;, 
\end{equation}
with the condition $c_{\{l_j\}}=
c_{\{-l_j\}}$ due to the fact that the cost is a real and even function. 
By performing the integrals in Eq. (\ref{avc2}) and exploiting the relation
$\int_0^{2\pi}d\phi e^{i(n-m)\phi}=\delta_{n,m}/2\pi$ we arrive at the 
following form of the average cost 
\begin{eqnarray}
\bar C=-c_0-\sum_{\{l_j\}\neq 0}
c_{\{l_j\}}\sum_{\{m_j-n_j\}=\{l_j\}}
\langle{\psi_0}\ket{\{n_j\}}\langle \{m_j\}\ket{\psi_0}\chi_{\{n_j\}\{m_j\}}\;,
\label{avcxi2}
\end{eqnarray}
where the expression $\{l_j\}\neq 0$ under the first summation symbol
means that the sum contains all the values of the indexes $l_j$ apart from 
the case where they are all zero (this contribution corresponds to the
term $c_0$ in Eq. (\ref{avcxi2})) and the expression
$\{m_j-n_j\}=\{l_j\}$ under the second summation 
means that the equality $m_j-n_j=l_j$ must hold for all values of the index 
$j$.

Let us now consider the following inequality
\begin{eqnarray}
{\mbox{sign}}(c_{\{l_j\}})\sum_{\{m_j-n_j\}=\{l_j\}}
\langle{\psi_0}\ket{\{n_j\}}\langle \{m_j\}\ket{\psi_0}
\chi_{\{n_j\}\{m_j\}}
\leq
\sum_{\{m_j-n_j\}=\{l_j\}}
|\langle{\psi_0}\ket{\{n_j\}}||\langle \{m_j\}\ket{\psi_0}|\;,
\label{inpom}
\end{eqnarray}
where we remind that we chose $\langle \{n_j\}\ket{\psi_0}>0$.
The above relation becomes an equality if the conditions
\begin{eqnarray}
\chi_{\{n_j\}\{m_j\}}={\mbox{sign}}(c_{\{m_j-n_j\}}) 
\label{rel}
\end{eqnarray}
can be fulfilled. In this case the minimum cost  takes the simple form
\begin{eqnarray}
\bar C=-c_0-\sum_{\{l_j\}\neq 0}
c_{\{l_j\}}\sum_{\{m_j-n_j\}=\{l_j\}}
|\langle{\psi_0}\ket{\{n_j\}}||\langle \{m_j\}\ket{\psi_0}|\;.
\label{minc}
\end{eqnarray}
Notice however that the positivity
of $\chi_\parallel$ is not generally guaranteed for any set of
values of ${\mbox{sign}}(c_{\{l_j\}})$. 

Let us now define a general class of cost functions, that extends the one
considered by Holevo \cite{Holevo}, with 
\begin{eqnarray}
c_{\{l_j\}}\geq 0, \forall \{l_j\}\neq 0\;.
\label{class}
\end{eqnarray}
For this class the conditions (\ref{rel}) are trivially satisfied 
for all the sets of values ${\{n_j\}}$ and the optimal POVM takes the 
explicit form
\begin{eqnarray}
d\mu_\parallel(\{\phi_j\})&=&\frac{d\phi_1}{2\pi}...\frac{d\phi_M}{2\pi} 
|e(\{\phi_j\})\rangle \langle
e(\{\phi_j\})|\label{pm2}\;,
\label{pomopt}
\end{eqnarray}
where the vectors $|e(\{\phi_j\})\rangle$ are defined as
\begin{eqnarray}
|e(\{\phi_j\})\rangle =\sum_{\{n_j\}} e^{i\sum_j n_j\phi_j}|\{n_j\}\rangle \;.
\label{sg}
\end{eqnarray}

In the following sections we will illustrate more explicitly the results
presented here by considering specific examples.

\section{Double phase estimation for qutrits}
\label{s:qutrits}

As a simple application of the concepts presented above, let us consider
the optimal double phase estimation for $N$ identical three-dimensional 
quantum systems (qutrits) all in the state
\begin{eqnarray}
|\psi(\phi,\theta)\rangle =\frac{1}{\sqrt 3}(\ket{0}+e^{i\phi}\ket{1}
+e^{i\theta}\ket{2})\;,
\label{qutrits}
\end{eqnarray}
where $\{\ket{0},\ket{1},\ket{2}\}$ represents a basis for the qutrit.
In the language of the previous section we identify 
$\phi_1=\phi$, $\phi_2=\theta$,
$\hat H_1=\ket{1}\bra{1}$, $\hat H_2=\ket{2}\bra{2}$ and
$\ket{\psi_0}=(\ket{0}+\ket{1}+\ket{2})/\sqrt{3}$ for each qutrit.
For the composite system of $N$ qutrits we have
$\hat H_1=\sum_{k=1}^N\ket{1}\bra{1}_k$ and $\hat H_2=\sum_{k=1}^N
\ket{2}\bra{2}_k$, where $\ket{j}\bra{j}_k$ denotes the projection operator 
onto the state $\ket{j}$ of the $k$-th qutrit. The operators $\hat H_1$
and $\hat H_2$ commute, and they are diagonalised in the basis 
$\ket{N-n_1-n_2,n_1,n_2}_\nu$ of the states where $N-n_1-n_2$ qutrits 
are in the state $\ket{0}$, $n_1$ in the state $\ket{1}$ and $n_2$ in the state
$\ket{2}$. The symbol 
$\nu$ represents the degeneracy index of the corresponding subspace,
and in particular it ranges from 1 to $N!/(N-n_1-n_2)!n_1!n_2!$.
Since the state of the $N$ qutrits is symmetric under any permutation 
performed on the qutrits, the states $\ket{\{n_j\}}$ defined in the
previous section by the projection method correspond in this case to
the symmetric normalised states of the $N$ qutrits, 
which we will simply denote as $\ket{N-n_1-n_2,n_1,n_2}_s$ (such a state is
an equally weighted superposition of $N!/(N-n_1-n_2)!n_1!n_2!$ components 
corresponding to all the possible permutations of states with $N-n_1-n_2$ 
qutrits in the state $\ket{0}$, $n_1$ in the state $\ket{1}$ and $n_2$ in the 
state $\ket{2}$).

In this case the optimal POVM (\ref{pomopt}) 
for the cost functions of the generalised Holevo form
(\ref{class}) takes the form
\begin{eqnarray}
d\mu(\phi,\theta)\equiv\frac{d\phi}{2\pi}\frac{d\theta}{2\pi}
\ket{e(\phi,\theta)}\bra{e(\phi,\theta)}\;,
\label{3POVM}
\end{eqnarray}
where $\ket{e(\phi,\theta)}=\sum_{n_1=0}^N\sum_{n_2=0}^{N-n_1}
e^{i(n_1\phi+n_2\theta)}\ket{N-n_1-n_2,n_1,n_2}_s$.

Let us now compute the fidelity of the optimal 
double phase estimation procedure.
As a cost function we can choose for example $1-F$, where $F$ is
the fidelity of the estimated state
$|\psi(\bar\phi,\bar\theta)\rangle$ with respect to the true state 
$|\psi(\phi,\theta)\rangle$. This cost belongs to the class (\ref{class}), 
and therefore the corresponding optimal POVM is the one written above.
This choice of the cost function is particularly interesting because the
fidelity is the figure of merit usually adopted to describe other processes
in quantum information theory, such as for instance cloning transformations, 
and therefore it allows a direct comparison of the efficiency of optimal
phase estimation with other procedures.

By the covariance of the procedure we can write the fidelity as
\begin{eqnarray}
F(\phi,\psi)=|\bra{\psi_0}\psi(\phi,\theta)\rangle|^2=
\frac{1}{9}[3+2\cos\phi+2\cos\psi+2\cos(\phi-\psi)]\;,
\label{F}
\end{eqnarray}
where 
\begin{eqnarray}
\ket{\psi_0}=\frac{1}{\sqrt{3^N}}\sum_{j=0}^N\sum_{k=0}^{N-j}\sqrt{M(N,j,k)}
\ket{N-j-k,j,k}_s\;.
\end{eqnarray}
In the above equation we have defined $M(N,j,k)\equiv\frac{N!}{(N-j-k)!j!k!}$.

The average fidelity $\bar F$ of the procedure is then given by
\begin{eqnarray}
\bar{F}&\equiv&\int F(\phi,\psi){\mbox{Tr}}[\rho_0 d\mu(\phi,\psi)]
\nonumber\\
&=&\frac{1}{9}\int_0^{2\pi}\frac{d\phi}{2\pi}
\int_0^{2\pi}\frac{d\psi}{2\pi}[3+2\cos\phi+2\cos\psi+2\cos(\phi-\psi)]
|\bra{\psi_0}\psi(\phi,\theta)\rangle|^2\;.
\label{barF}
\end{eqnarray}
By performing the integration in Eq. (\ref{barF}) we have
\begin{eqnarray}
\bar{F}&=&\frac{1}{3}+\frac{1}{3^{N+2}}\sum_{j,p=0}^N\sum_{k=0}^{N-j}
\sum_{q=0}^{N-p}\sqrt{M(N,j,k)M(N,p,q)}\nonumber\\
&&\times\left[\delta_{k,q}(\delta_{j,p+1}+\delta_{j+1,p})
+\delta_{j,p}(\delta_{k,q+1}+\delta_{k+1,q})+
\delta_{j+1,p}\delta_{k,q+1}+\delta_{j,p+1}\delta_{k+1,q}\right]
\nonumber\\
&=&\frac{1}{3}+\frac{2}{3^{N+1}}\sum_{j=0}^{N-1}\sum_{k=0}^{N-j-1}
M(N,j,k)\sqrt{\frac{N-j-k}{j+1}}\;.
\label{barFf}
\end{eqnarray}

We want to point out 
that the fidelity (\ref{barFf}) corresponding to the optimal
double phase estimation for qutrits 
is always smaller than the one for equatorial qubits, given in Ref. \cite{dbe},
where a single phase is estimated.

We want also to stress that, as in the case of qubits, there is a relation
between optimal double phase estimation and optimal cloning for states of the
form (\ref{qutrits}). Actually, the fidelity of the optimal double phase
estimation 
(\ref{barFf}) for a single qutrit ($N=1$) coincides with the cloning fidelity 
for the optimal $1\to M$ cloning transformations, 
that take a single input equatorial qutrit
and produce $M$ output copies \cite{opt_pcc},
in the limit of an infinite number of output copies, i.e. $M\to\infty$.
Moreover, this result is consistent also with the relation between optimal 
state estimation \cite{ddim} and optimal cloning \cite{werner} for input 
qutrits whose state is completely unknown (not restricted to the form
(\ref{qutrits})).

We want to point out that other figure of merits could be considered in 
order to evaluate the efficiency of the phase estimation procedure. For 
example, a mean periodic ``variance'' $V(\phi,\theta)=2(\sin^2 \phi/2+
\sin^2\theta/2)$ could be considered as a cost function. In this case the 
cost function still belongs to the class (\ref{class}), and therefore, by 
explicitly calculating the average variance in a similar way as for the 
average fidelity, we arrive at the form

\begin{eqnarray}
\bar{V}\equiv\int V(\phi,\psi){\mbox{Tr}}[\rho_0 d\mu(\phi,\psi)]
=2-\frac{2}{3^{N}}\sum_{j=0}^{N-1}\sum_{k=0}^{N-j-1}
M(N,j,k)\sqrt{\frac{N-j-k}{j+1}}\;.
\label{barV}
\end{eqnarray}

\section{Multiple phase estimation for systems with arbitrary dimension}
\label{s:qudits}

In this  section we generalise the results derived above for qutrits to 
the case of multiple phase estimation for systems with arbitrary finite
dimension $d$ (qudits). We will consider the optimal multiple phase estimation 
for $N$ identical $d$-dimensional quantum systems all in the state
\begin{eqnarray}
|\psi(\{\phi_j\})\rangle =\frac{1}{\sqrt d}(\ket{0}+e^{i\phi_1}\ket{1}
+e^{i\phi_2}\ket{2}+...+e^{i\phi_{d-1}}\ket{d-1})\;,
\label{qudits}
\end{eqnarray}
where $\{\ket{0},\ket{1},\ket{2}
...\ket{d-1}\}$ represents a basis for each
system. 

In the language of section \ref{s:problem}, we are considering the estimation 
problem for $d-1$ phases corresponding to the operators 
$\hat H_j=\ket{j}\bra{j}$, $j=1,..,d-1$, and
$\ket{\psi_0}=(\ket{0}+\ket{1}+\ket{2}+...+\ket{d-1})/\sqrt{d}$ 
for each system.
For the composite system of $N$ qudits we have
$\hat H_j=\sum_{k=1}^N\ket{j}\bra{j}_k$, where as in the previous section
$\ket{j}\bra{j}_k$ 
denotes the projection operator onto the state $\ket{j}$ of the $k$-th qudit. 
The operators $\hat H_j$ commute, and they are diagonalised in the basis 
$\ket{n_0,n_1,n_2,...n_{d-1}}_\nu$ of the states where 
$n_0$ qudits are in the state $\ket{0}$, $n_1$ in the state $\ket{1}$ and so 
on, with $\sum_{j=0}^{d-1} n_j=N$.
As in the case of qutrits, $\nu$ represents the degeneracy index of the 
corresponding subspace, and in particular it ranges from 1 to the multinomial 
$N!/(N-n_1-n_2-...n_{d-1})!n_1!n_2!...n_{d-1}!$.
Analogously to the case of qutrits,
the POVM is optimised by choosing the symmetric normalised states of the 
$N$ qudits, 
which we will simply denote as $\ket{n_0,n_1,n_2,...n_{d-1}}_s$.

The optimal POVM for the cost functions of the generalised Holevo form
(\ref{class}) is given by (\ref{pomopt}), with $M=d-1$ and 

\begin{eqnarray}
\ket{e(\{\phi_j\})}=\sum_{\{n_j\}}
e^{i\sum_{j=1}^{d-1}n_j\phi_j}\ket{n_0,n_1,n_2,...n_{d-1}}_s\;.
\end{eqnarray}
In the above equation the sum over $\{n_j\}$ means that the variables $n_j$ 
take all the possible non negative values compatible with the constraint 
$\sum_{j=0}^{d-1} n_j=N$.

Let us now compute the fidelity of the optimal multiple phase estimation 
procedure derived above.
As in the case of qutrits we choose a cost function of the form
$1-F$, where $F$ is the fidelity of the estimated state
$|e(\{\bar\phi_j\})\rangle$ with respect to the true state 
$|e(\{\phi_j\})\rangle$. This cost belongs to the class (\ref{class}), 
and therefore the corresponding optimal POVM is the one mentioned above.
By the covariance of the procedure we can write the fidelity as
\begin{eqnarray}
F(\{\phi_j\})=|\bra{\psi_0}\psi(\{\phi_j\})\rangle|^2=
\frac{1}{d^2}[d+2\sum_{j=1}^{d-1}\cos\phi_j+2\sum_{j>k}
\cos(\phi_j-\phi_k)]\;
\label{Fd}
\end{eqnarray}
where 
\begin{eqnarray}
\ket{\psi_0}=\frac{1}{\sqrt{d^N}}\sum_{\{n_j\}}
\sqrt{\frac{N!}{n_0!n_1!n_2!...n_{d-1}!}}
\ket{n_0,n_1,n_2,...n_{d-1}}_s\;.
\end{eqnarray}
The average fidelity $\bar F$ of the procedure is now given by
\begin{eqnarray}
\bar{F}
=\frac{1}{d^2}\int_0^{2\pi}\frac{d\phi_0}{2\pi}...
\int_0^{2\pi}\frac{d\phi_{d-1}}{2\pi}F(\{\phi_j\})
|\bra{\psi_0}\psi(\{\phi_j\})\rangle|^2\;.
\label{barFd}
\end{eqnarray}
By performing the integrations in Eq. (\ref{barFd}) we have
\begin{eqnarray}
\bar{F}=&&\frac{1}{d}+\frac{d-1}{d^{N+1}}\sum_{n_1=0}^{N-1}
\sum_{n_2=0}^{N-n_1-1}
...\sum_{n_{d-1}=0}^{N-n_1-n_2-...-1}
\frac{N!}{(N-n_1-n_2-...n_{d-1})!n_1!n_2!...n_{d-1}!}\nonumber\\
&&\sqrt{\frac{N-n_1-n_2-...n_{d-1}}{n_1+1}}\;.
\label{barFfd}
\end{eqnarray}

Notice that the fidelity decreases as a function of the dimension $d$.
For example, in the case of multiple phase estimation on the state 
of a single qudit it takes the simple form

\begin{eqnarray}
\bar{F_1}=\frac{2d-1}{d^2}\;.
\label{FdN=1}
\end{eqnarray}

We also want to point out 
that the above fidelity is larger than the fidelity of estimation of a single
qudit in a completely unknown pure state (not restricted to be of the form
(\ref{qudits})). Actually, the fidelity of such a universal procedure, which
we will call $\bar{F}_{univ,1}$, is given by \cite{ddim}

\begin{eqnarray}
\bar{F}_{univ,1}=\frac{2}{d+2}\;.
\label{Fest}
\end{eqnarray}


\section{Conclusions}
\label{s:conc}

In this paper we have addressed the problem of simultaneous estimation
of several phase shifts induced by a unitary transformation acting
on a quantum system. We have derived in a general way 
the optimal estimation procedure for an 
arbitrary number of phase shifts and for a wide class of cost functions.
We have then specialised the results obtained to the case of ``equatorial'' 
qutrits and then generalised them to the case of quantum system with 
arbitrary finite dimension. An interesting result that was
emphasised in this work is the connection between optimal double phase 
estimation 
and optimal double phase covariant cloning for qutrits. We expect also that 
such a connection is valid in arbitrary finite dimension.

Before closing the paper, we want to point out that the 
scenario of simultaneous estimation of several phases considered in this 
work may be exploited to design schemes where several
variables are encoded into phases in the same quantum states and 
in this way the efficiency of quantum information processing tasks may be
improved. 

\section*{Acknowledgements}

This work has been supported in part by the EC programs
ATESIT (Contract No. IST-2000-29681) and 
QUPRODIS (Contract No. IST-2002-38877).


\begin{thebibliography}{99}
\bibitem{nota} The case of non-commuting operators is highly non trivial
and is not considered in this paper.
\end{thebibliography}
\end{document}